\documentclass[english,prl,showpacs,reprint,superscriptaddress]{revtex4-1}
\usepackage[latin9]{inputenc}
\usepackage{color}
\usepackage{babel, array, multirow}
\usepackage{verbatim}
\usepackage{bm}
\usepackage{amsmath}
\usepackage{amssymb}
\usepackage{graphicx}
\usepackage{gensymb}
\usepackage[unicode=true,pdfusetitle,bookmarks=true,bookmarksnumbered=false,bookmarksopen=false,breaklinks=false,pdfborder={0 0 0},pdfborderstyle={},backref=false,colorlinks=true]{hyperref}
\hypersetup{citecolor=blue}
\makeatletter
\usepackage{epstopdf}
\usepackage{amsfonts}
\usepackage{mathrsfs}\usepackage{bm}\usepackage{appendix}\usepackage{color}
\usepackage{txfonts}
\usepackage{float}
\usepackage{epstopdf}\usepackage{ulem}
\usepackage{textcomp}

\newcommand{\Rmnum}[1]{\expandafter\@slowromancap\romannumeral #1@}
\usepackage{CJK}
\usepackage{colortbl}
\makeatother

\begin{document}

\title{Observation of fourfold Dirac nodal line semimetal and its unconventional surface responses in sonic crystals}
\author{Chang-Yin Ji}
\altaffiliation{These two authors contribute equally to this work.}
\affiliation{Key Lab of advanced optoelectronic quantum architecture and measurement (MOE), Beijing Key Lab of Nanophotonics $\&$ Ultrafine Optoelectronic Systems, and School of Physics, Beijing Institute of Technology, Beijing 100081, China}
\affiliation{School of Integrated Circuits and Electronics, MIIT Key Laboratory for Low-Dimensional Quantum Structure and Devices, Beijing Institute of Technology,Beijing 100081, China}

\author{Xiao-Ping Li}
\altaffiliation{These two authors contribute equally to this work.}
\affiliation{School of Physical Science and Technology, Inner Mongolia University, Hohhot 010021, China}

\author{Zheng Tang}
\affiliation{Key Lab of advanced optoelectronic quantum architecture and measurement (MOE), Beijing Key Lab of Nanophotonics $\&$ Ultrafine Optoelectronic Systems, and School of Physics, Beijing Institute of Technology, Beijing 100081, China}

\author{Di Zhou}
\affiliation{Key Lab of advanced optoelectronic quantum architecture and measurement (MOE), Beijing Key Lab of Nanophotonics $\&$ Ultrafine Optoelectronic Systems, and School of Physics, Beijing Institute of Technology, Beijing 100081, China}

\author{Yeliang Wang}
\affiliation{School of Integrated Circuits and Electronics, MIIT Key Laboratory for Low-Dimensional Quantum Structure and Devices, Beijing Institute of Technology,Beijing 100081, China}

\author{Feng Li}
\email[]{phlifeng@bit.edu.cn}
\affiliation{Key Lab of advanced optoelectronic quantum architecture and measurement (MOE), Beijing Key Lab of Nanophotonics $\&$ Ultrafine Optoelectronic Systems, and School of Physics, Beijing Institute of Technology, Beijing 100081, China}

\author{Jiafang Li}
\email{jiafangli@bit.edu.cn}
\affiliation{Key Lab of advanced optoelectronic quantum architecture and measurement (MOE), Beijing Key Lab of Nanophotonics $\&$ Ultrafine Optoelectronic Systems, and School of Physics, Beijing Institute of Technology, Beijing 100081, China}

\author{Yugui Yao}
\affiliation{Key Lab of advanced optoelectronic quantum architecture and measurement (MOE), Beijing Key Lab of Nanophotonics $\&$ Ultrafine Optoelectronic Systems, and School of Physics, Beijing Institute of Technology, Beijing 100081, China}

\begin{abstract}
Three-dimensional nodal line semimetals (NLSMs) provide remarkable importance for both enrich topological physics and wave management. However, NLSMs realized in acoustic systems are twofold bands degenerate, which are called Weyl NLSMs. Here, we first report on the experimental observation of novel Dirac NLSMs with fourfold degenerate in sonic crystals. We reveal that the topological properties of the Dirac NLSMs are entirely \textit{different} than that of the conventional Weyl NLSMs. The Berry phase related to the Dirac nodal line (DNL) is 2$\pi$, which results in the surface responses of the Dirac NLSMs with two radically different situations: a torus surface state occupying the entire surface Brillouin zone (SBZ) and without any surface state in the SBZ. We further reveal that topological surface arcs caused by DNL can change from open to closed contours. The findings of Dirac NLSMs and their unique surface response may provoke exciting frontiers for flexible manipulation of acoustic surface waves.
\end{abstract}

\maketitle
\textcolor{blue}{\textit{Introduction}}. Three-dimensional (3D) topological semimetals (TSMs) have been one of the most flourishing research fields since the discovery of Weyl fermion in the solid materials \cite{wan2011topological}. Compared with topological insulators, 3D TSMs open another new research perspective to explore plentiful exotic topological phenomena and electronic transport \cite{armitage2018weyl,hu2019transport,lv2021experimental}, including the topologically protected surface states, chiral anomaly \cite{zyuzin2012topological}, planar Hall effect \cite{kumar2018planar, ma2019planar}, intrinsic anomalous Hall effect \cite{zyuzin2016intrinsic, thakur2020intrinsic}, quantized circulation of anomalous shift \cite{liu2020quantized} and so on.  According to the dimension of the band degeneracies of 3D TSMs \cite{park2022nodal}, we can classify them into zero-dimensional (0D)
nodal points, one-dimensional nodal lines (NLs) and two-dimensional nodal surface semimetals \cite{yang2019observation,xiao2020experimental}. Since the topological band theory in condensed matter physics is fundamental and universal to classical wave systems, the concepts of TSMs have been extended from quantum systems into photonic and acoustic systems in the last few years. Experimental observations have shown that the unprecedented flexibility in design afforded by macroscopic meta-atoms has led to the discovery of classical waves analog Weyl semimetals and Dirac semimetals \cite{he2018topological,li2018weyl,he2020observation,xia2019observation, cheng2020discovering,cai2020symmetry,xie2020dirac}. The experimental results provide solid evidence for Weyl and Dirac points endowed topological properties, such as the fermi arcs arising from Weyl point \cite{he2018topological,li2018weyl,he2020observation}, gapless helicoid surface states arising from Dirac point \cite {cheng2020discovering,cai2020symmetry} and the high order hinge states connecting the projected Dirac or Weyl points \cite{qiu2021higher,luo2021observation,wei2021higher}.

\begin{figure}
\centering{}\includegraphics[width=0.45\textwidth]{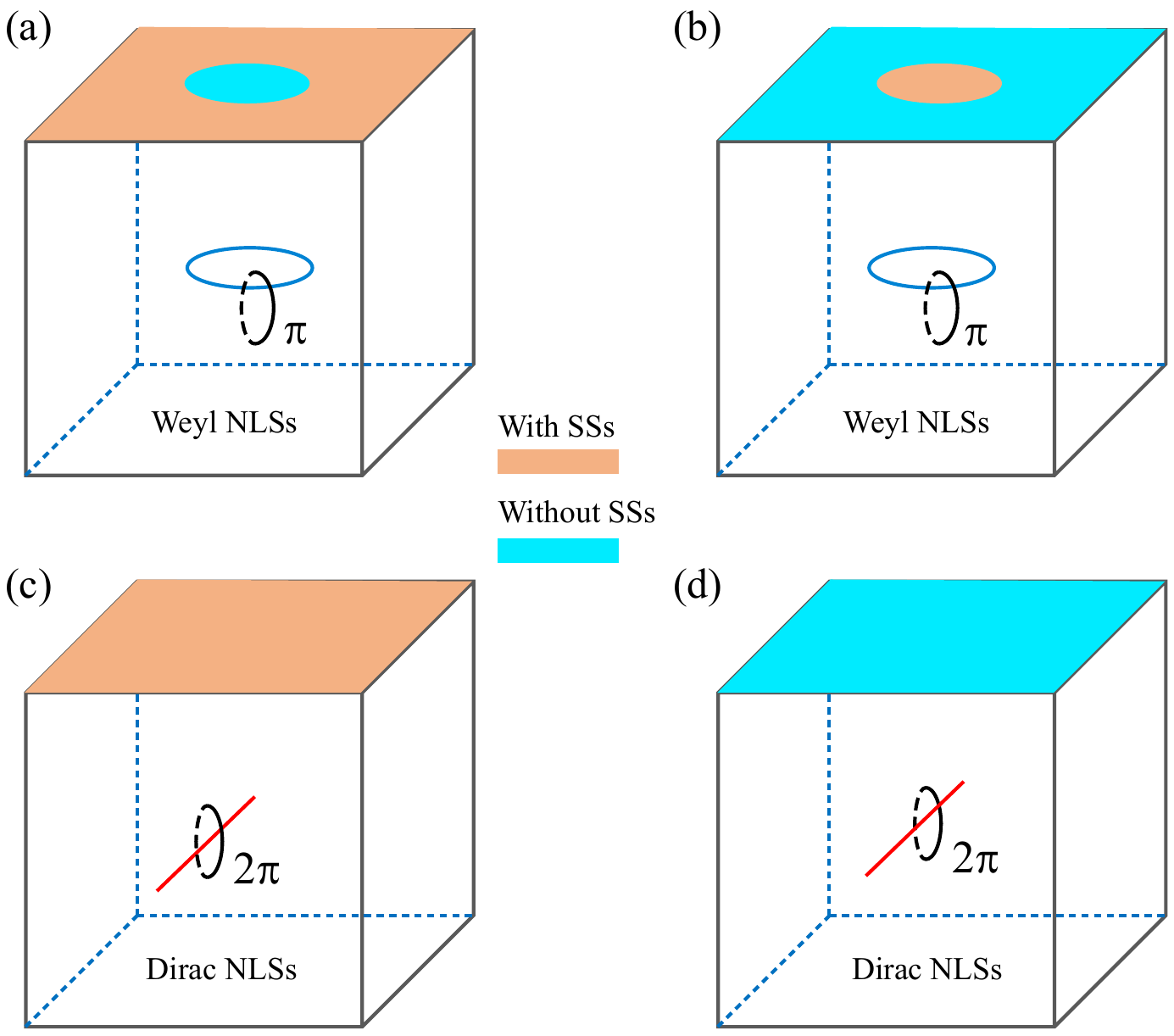}
\caption{Schematic illustration the Berry phase and the related surface states (SSs) of the nodal line semimetals (NLSMs). (a, b) Weyl NLSMs with twofold degeneracy along line in the momentum space. The Berry phase of the Weyl NLSMs is $\pi$, which leads to drumhead SSs that occupy the region outside nodal ring in (a) or inside nodal ring in (b). (c, d) Dirac NLSMs with fourfold degeneracy along line in the momentum space. The Berry phase of the Dirac NLSMs is 2$\pi$. The Dirac NLSs have two distinct surface responses: a torus SSs occupying the entire surface Brillouin zone in (c) or without any SSs in (d).}
\label{Fig-1}
\end{figure}

\begin{figure*}
\centering{}\includegraphics[width=0.9\textwidth]{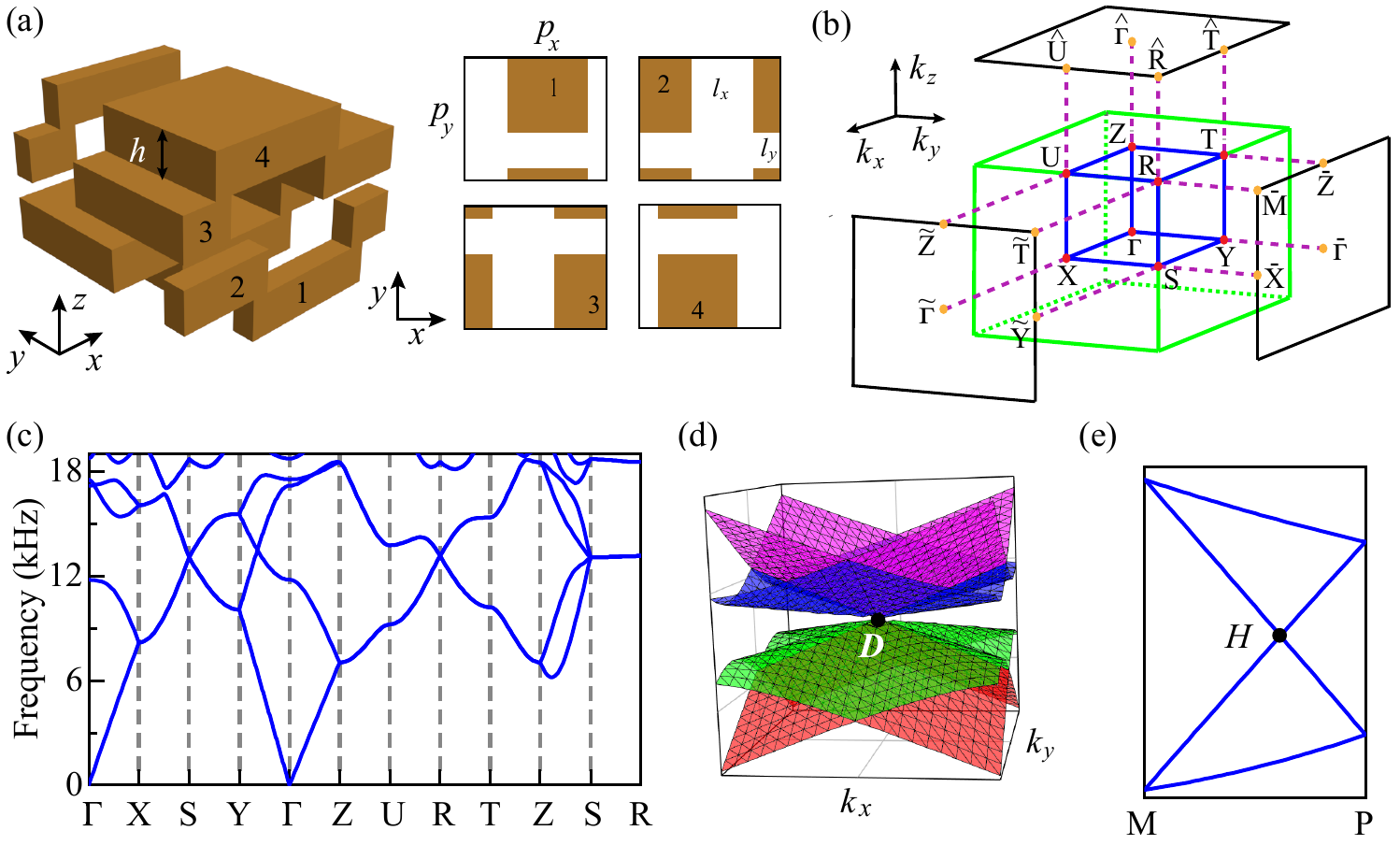}
\caption{(a) Unit-cell structure of the designed layer acoustic Dirac NLSMs. The lattice periods along the $x$, $y$ and $z$ directions are $p_x=1.84$ cm, $p_y=1.6$ cm, and $p_z=1.52$ cm, respectively. The cell contains four layers, where the first and second layers are connected to each other, and the third and fourth layers are connected to each other. The center positions of the first to fourth layers along the $z$ direction are $z_0$, $p_z/2-z_0$, $p_z/2+z_0$ and $p_z-z_0$, respectively. $z_0$=0.23 cm. The thickness of each layer is both $h=0.3$ cm. The width of the air slit in each layer along the $x$ and $y$ directions are $l_x=0.8$ cm, $l_y=0.46$ cm, respectively. (b) Bulk Brillouin zone (BZ) and (100), (010)and (001) surface BZ. (c) Bulk band structure of the Dirac-Weyl NLSs in (a). (d) A Dirac shaped dispersion around a generic $\bm k$ point along SR. (e) A hourglass shaped dispersion along a generic path of MP. Points M and P are arbitrary points on paths XS and SY respectively}
\label{Fig-2}
\end{figure*}

Compared to the widely studied nodal points semimetals in classical systems, the one-dimensional nodal lines (NLs) are still poorly studied \cite{park2022nodal,deng2019nodal,qiu2019straight}. The configurations and topological properties of nodal lines are more diverse compared to nodal point semimetals. According to the shape of one-dimensional NLs and the connectivities between NLs, they can be classified into different types, such as nodal curve, nodal ring, nodal chain, nodal link, nodal knot and so on \cite{park2022nodal}.
According to the number of band degeneracy, NLs can be divided into Weyl NLs with twofold line degeneracy and Dirac NLs with fourfold line degeneracy \cite{li2021double,yu2022encyclopedia}. The Berry phase of the Weyl NLs is $\pi$, as shown in Fig.~\ref{Fig-1}(a, b). This results in the drumhead surface states (SSs), which can only cover a small region of Brillouin zone (BZ), for example the region outside or inside nodal ring, as shown in Fig.~\ref{Fig-1}(a, b). In current acoustic systems, only Weyl Nodal Line Semimetals (NLSMs) have been realized, while Dirac Nodal Line Semimetals are still lacking.

Here, we theoretically propose and experimentally realize the Dirac NLSMs to fill this void. One step further, we demonstrate that the topological properties and surface responses caused by Dirac nodal line (DNL) are distinct from that of the Weyl nodal line (WNL). The Berry phase of the DNL is 2$\pi$, as shown in Fig.~\ref{Fig-1}(c, d). Such 2$\pi$ Berry phase leads to intriguing surface responses which are entirely different from that of WNL. As shown in Fig.~\ref{Fig-1}(c), the SSs can occupy the entire surface Brillouin zone forming the torus SSs. The other scenario is that the whole BZ has no any SSs, as shown in Fig.~\ref{Fig-1}(d). Compared to the situation of the surface responses induced by Weyl NLSMs, the Dirac NLSMs have unconventional surface responses.

\textcolor{blue}{\textit{Theoretical Design}}.
The designed acoustic Dirac NLSM is shown in Fig.~\ref{Fig-2}(a), which is an orthorhombic structure and belongs to space group $Pnma$ (No. 62). The lattice periods along the $x$, $y$ and $z$ directions are $p_x=1.84$ cm, $p_y=1.6$ cm, and $p_z=1.52$ cm, respectively. There are four symmetry operators for the Diracl NLSMs in Fig.~\ref{Fig-2}(a), two screw rotations $\tilde{C}_{2z}:(x,y,z)\rightarrow (-x+p_x/2,-y,z+p_z/2)$ and $\tilde{C}_{2y}:(x,y,z)\rightarrow (-x,y+p_y/2,-z)$, and spatial inversion ${\cal P}:(x,y,z)\rightarrow (-x,-y,-z)$, and time-reversal symmetry $\cal T$. The bulk and surface Brillouin zone are shown in Fig.~\ref{Fig-2}(b). The Fig.~\ref{Fig-2}(c) is the band structure of the designed acoustic crystal in Fig.~\ref{Fig-2}(a). It can be seen that there exists fourfold line degeneracy along the high symmetry line of SR, which confirms that Dirac nodal line is present. The Dirac nodal line is a open straight nodal line and located in a fairly wide and clean frequency range, which can serve as an ideal DNL.

We further use symmetry analysis to illustrate why the DNL appears at the hinge between the $k_x=\pi/p_x$ and $k_y=\pi/p_y$ planes. For an arbitrary $D$ point located on the SR line, its symmetry operators can be expressed as, $\tilde{C}_{2z}$, $\tilde{M}_{y}:(x,y,z)\rightarrow (x,-y+p_y/2,z)$ and a combined operator ${\cal A}=\tilde{C}_{2y}\cal T$. They meet the following form,
\begin{eqnarray}
  \tilde{C}_{2z}^2 &=&1,\quad \tilde{M}_{y}^2 =1,\quad {\cal A}^2 =-1.
\end{eqnarray}

\begin{figure*}
\centering{}\includegraphics[width=0.8\textwidth]{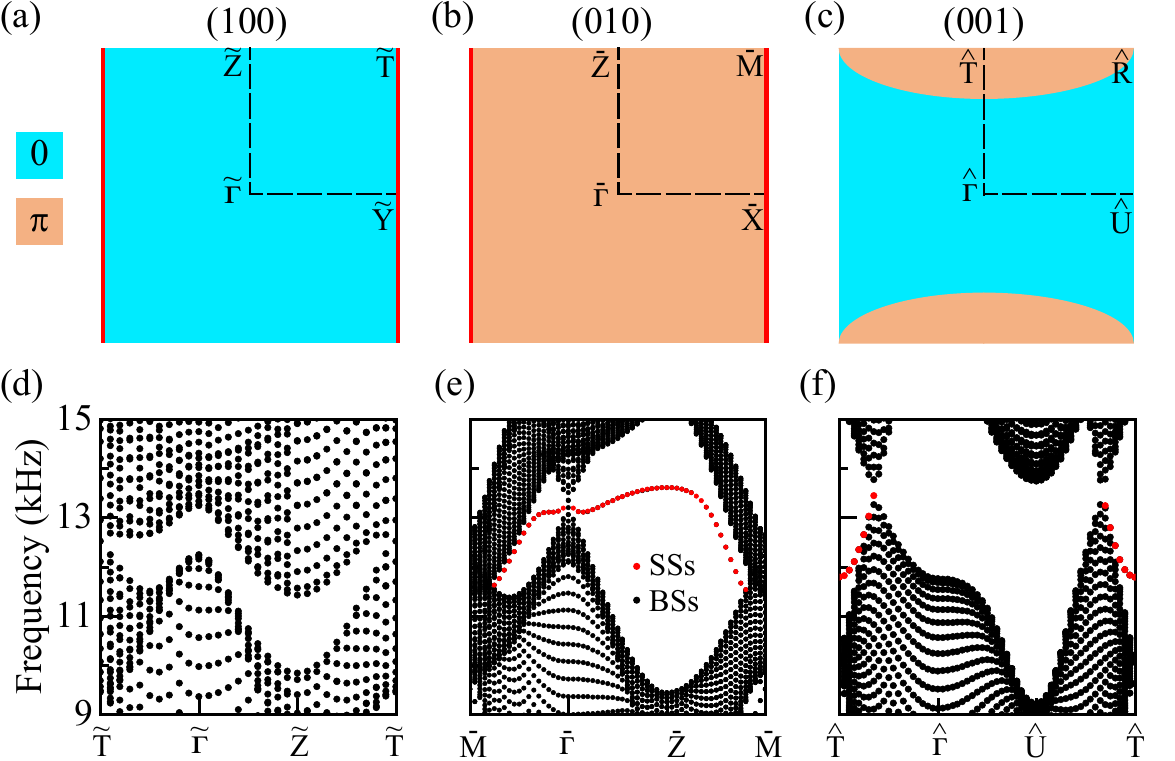}
\caption{(a-c) The Zak phase distributios on the surfaces of (100),(010) and (001). The cyan and orange regions indicate the Zak phases 0 and $\pi$, respectively. The red lines in (a) and (b) show the position of Dirac nodal line. (d-f) Simulated surface spectra on (100), (010), (001) surfaces, respectively. The red and black dots represent the nontrivial surface states (SSs) and bulk states (BSs), respectively. }
\label{Fig-3}
\end{figure*}

The combination of $\tilde{M}_{y}$ and $\tilde{C}_{2z}$ yields,
\begin{eqnarray}
  \tilde{M}_{y}\tilde{C}_{2z}:(x,y,z)\rightarrow (-x+p_x/2,y+p_y/2,z+p_z/2), \nonumber \\
  \tilde{C}_{2z}\tilde{M}_{y}:(x,y,z)\rightarrow (-x+p_x/2,y-p_y/2,z+p_z/2).
\end{eqnarray}
Let $|\varphi\rangle$ is an eigenstate of $\tilde{C}_{2z}$ at the $D$ point, one can get $\tilde{M}_{y}\tilde{C}_{2z}|\varphi\rangle=e^{ik_yp_y/2}[e^{i(k_xp_x/2+k_zp_z/2)}M_x|\varphi \rangle]$ and $\tilde{C}_{2z}\tilde{M}_{y}|\varphi\rangle=e^{-ik_yp_y/2}[e^{i(k_xp_x/2+k_zp_z/2)}M_x|\varphi\rangle]$. Due to $k_y=\pi/p_y$ at the $D$ point, one can get $\tilde{M}_{y}\tilde{C}_{2z}=-\tilde{C}_{2z}\tilde{M}_{y}$.
Similar analysis can be applied to the combination operates of $\tilde{C}_{2z}{\cal A}$ and $\tilde{M}_{y}{\cal A}$, one can get $\tilde{C}_{2z}{\cal A} ={\cal A}\tilde{C}_{2z}$ and $\tilde{M}_{y}{\cal A} =-{\cal A}\tilde{M}_{y}$. For the convenience of discussion, we take $|\varphi\rangle$ as the eigenstate of the eigenvalue
of 1 of the $\tilde{C}_{2z}$. Then, we can have
\begin{eqnarray}
  \tilde{C}_{2z}\tilde{M}_{y}|\varphi\rangle=-\tilde{M}_{y}|\varphi\rangle,\quad \langle \varphi|\tilde{M}_{y}|\varphi\rangle=0.
\end{eqnarray}
Thus, the eigenstates of $|\varphi\rangle$ and $\tilde{M}_{y}|\varphi\rangle$ are a pair of degenerate states. Since there exists antiunitary operator $\cal A$ on the high symmetry line of $SR$, the $|\varphi\rangle$ and $\tilde{M}_{y}|\varphi\rangle$ have Kramers-like counterparts of ${\cal A}|\varphi\rangle$ and ${\cal A}\tilde{M}_{y}|\varphi\rangle$, respectively. In addition, since $\tilde{C}_{2z}$ and ${\cal A}$ are commutative, the states and their Kramers-like counterparts are linearly independent. Thus, there must exist four degenerate states of $|\varphi\rangle$, $\tilde{M}_{y}|\varphi\rangle$, ${\cal A}|\varphi\rangle$ and ${\cal A}\tilde{M}_{y}|\varphi\rangle$ at the any $\bm k$ point on the high symmetry line of SR. The Fig.~\ref{Fig-2}(d) is the band dispersion around a certain $\bm k$ point along SR. It can be seen that Dirac point splits at ordinary $\bm k$ point, forming a Dirac shaped dispersion.

With $\bm k\cdot\bm p$ perturbation method, we also derive a four-band Hamiltonian to gain a deeper insight into the physics around the DNL,
\begin{equation}\label{h_d}
  {\cal H}_{DNL}=a_0+a_1k_z+\left(
                              \begin{array}{cc}
                                h_0 & h_1 \\
                                h_1^{\dag} & h_0 \\
                              \end{array}
                            \right),
\end{equation}
where $h_0=v_x\delta k_x \sigma _x+v_y\delta k_y \sigma _y$ and $h_1=b_1\delta k_x \sigma _y+b_2\delta k_y \sigma _x$. Here, $a_i$ and $b_i$ are the real and complex numbers, respectively. $(\delta k_x, \delta k_y)$ denotes the momentum deviation from $(\pi/p_x, \pi/p_y)$. $\sigma _{i}$ is the Pauli matrice. $h_0$ describes the Dirac points in the $k_x$-$k_y$ plane, whose Berry phase is $\pi$. $v_x$ and $v_y$ are the Dirac velocity along the $x$ and $y$ directions, respectively. Fig.~\ref{Fig-2}(d) is the dispersion around the Dirac point. More interestingly, the Hamiltonian of Eq.~\ref{h_d} shows that the Berry phase of DNL is $2\pi$, which is very different from that of WNL.

Due to the non-symmorphic symmetry of the structure, it should be noticed that our designed acoustic crystal also has unrealized hourglass WNL in acoustic system. The Fig.~\ref{Fig-2}(e) is the hourglass-shaped dispersion along the patch of MP. The cross point $H$ must form a ring in the $k_x$-$k_y$ plane, which is guaranteed by the symmetry operates of $\tilde{M}_{z}=\tilde{C}_{2z}\cal P$ and ${\cal PT}$. The Berry phase of the hourglass WNL is the same as that of an ordinary nodal line, which is also $\pi$. Thus, our designed acoustic crystal provides an excellent platform to study the differences between WNL and DNL.

\begin{figure*}
\centering{}\includegraphics[width=0.9\textwidth]{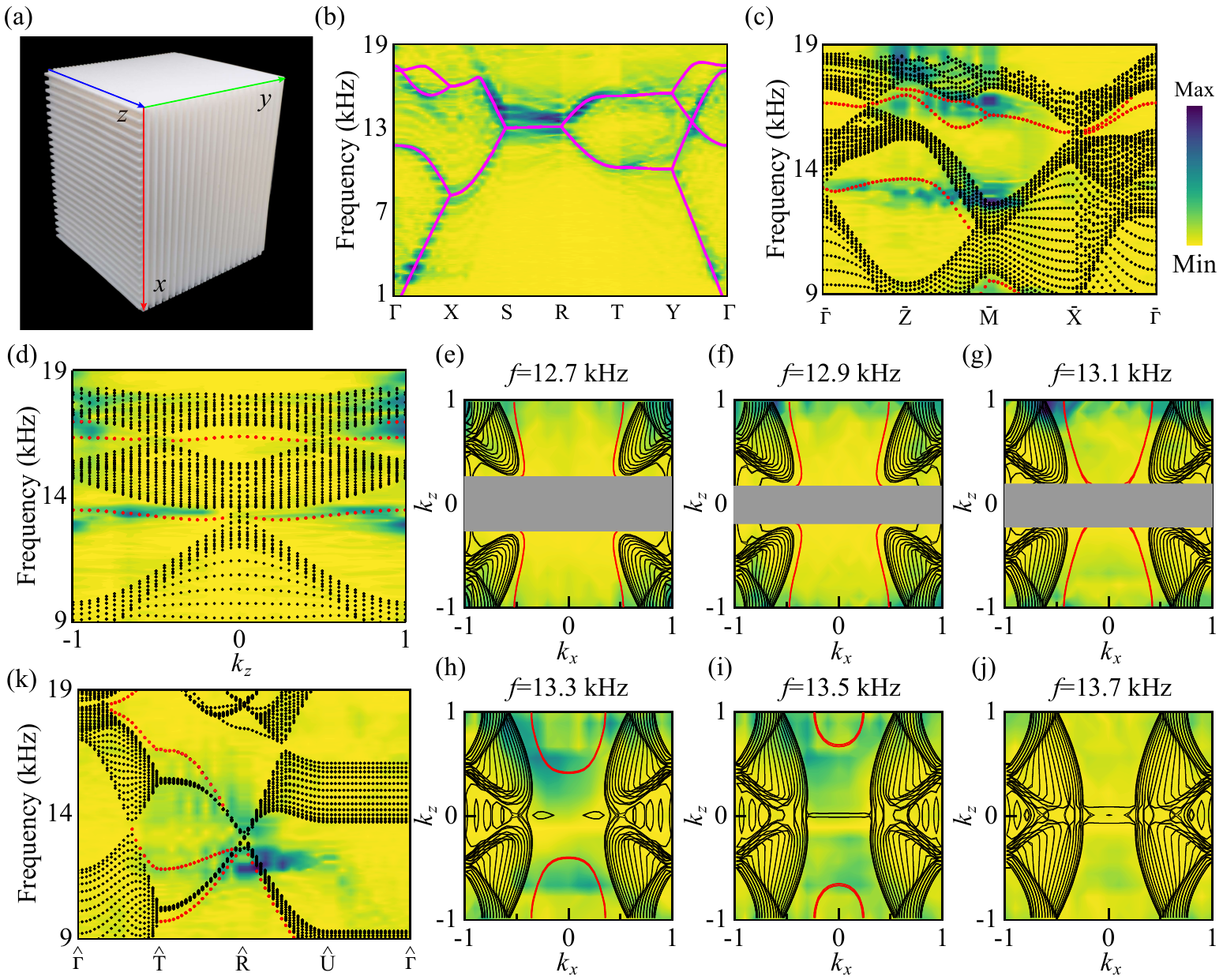}
\caption{(a) A photograph of the experimental sample fabricated by 3D printing. (b) The experimentally
measured (colors) bulk band structure along the high symmetry lines of the bulk BZ in Fig.~\ref{Fig-1}(b). The magenta solid lines are the simulated bulk band structure. (c) The experimentally measured surface dispersions of the (010) plane. (d) Measured dispersions with $k_x=0.3\pi/p_x$ for the (010) plane. (e-i) Measured isofrequency contours (coloured) of the (010) plane for different frequencies. The red curves show the topological surface arcs. (k) The experimentally measured surface dispersions of the (001) plane. The dot plot is the result of theoretical calculation. Red and black dots represent surface states and bulk states, respectively.}
\label{Fig-4}
\end{figure*}

An important feature of NLSs is the appearance of topologically protected SSs on their surfaces. To better reveal the surface response of the NLSMs, we first give the Zak phase distribution for the (100), (010) and (001) planes. The Zak phase is the Berry phase of a straight line perpendicular to the surface and passing through the bulk BZ. The existence of $\cal P$ requires the Zak phase to be quantized, whose values are 0 or $\pi$. The calculated Zak phase distributions is shown in Fig.~\ref{Fig-3}(a-c). It should be noticed that the (100) and (010) planes are parallel to the DNL, while the (001) plane is parallel to the WNL. The Fig.~\ref{Fig-3}(a) and Fig.~\ref{Fig-3}(b) show that the Zak Phase is equal to 0 and $\pi$ in the whole BZ of the (100) and (010) planes, respectively. This is caused by the $2\pi$ Berry phase of the DNL. For the (100) and (010) planes, the Zak phase of a straight line perpendicular to the surface can not be changed until the line passes through DNL. As it passes through the DNL, its Zak phase changes to $2\pi$. Since Zak phase is defined mod $2\pi$, the Zak phase of the entire BZ is the same, either 0 in Fig.~\ref{Fig-3}(a) or $\pi$ in Fig.~\ref{Fig-3}(b). The result in Fig.~\ref{Fig-3}(c) shows that only a small region of Zak phase is $\pi$, which results from the $\pi$ Zak phase of the hourglass WNL. Similarly, the Zak Phase of the straight line perpendicular to (001) plane will change $\pi$ once it passes through hourglass WNL. This necessarily results in the Zak phase of 0 in a portion region and the Zak phase of $\pi$ in the complementary region. The simulated surface spectra are shown in Fig.~\ref{Fig-3}(d-f). It can be seen that there are no any SSs in the whole BZ of (100) plane due to the Zak Phase is 0, as shown in Fig.~\ref{Fig-3}(d). However, the result in Fig.~\ref{Fig-3}(e) shows that the SSs cover the whole BZ, which is consistent with the Zak phase distribution of $\pi$. Since the two-dimensional BZ can be regarded as a torus, such surface state are called a novel torus surface state. The SSs in Fig.~\ref{Fig-3}(f) is the same as the usual Weyl NLSMs, featuring a drumhead-like SSs. Thus, our NLSMs can have three types of surface responses at the same time, namely, no any surface state, torus surface states and drumhead-like surface states. This rich surface response makes the Dirac NLSMs remarkably different from that of the widely studied Weyl NLSMs.

\textcolor{blue}{\textit{Experimental Observation}}. We further conduct a acoustic experiment to investigate the fascinating topological properties of the Dirac NLSMs. The sample is manufactured with 3D printing, as shown in Fig.\ref{Fig-4}(a). The sample contains $19\times 19 \times 19$ unit cells with the
size of $34.96\times 30.4 \times 28.88$ mm. To measure the bulk band structure, a tiny microphone is inserted into the sample as a sound source to excite the Bloch states. The frequency range of the source is 1 to 19 kHz. Then, we use a moving detector to record the spatial distributions of both the phase and amplitude of the Bloch states. Finally, the bulk band structure in the momentum space can be obtained by fourier transforming the measured Bloch states in real space. The measured bulk band structure along the high symmetry line of BZ is shown in Fig.\ref{Fig-4}(b), which is consistent with the theoretical calculation result [see the magenta solid lines in Fig.\ref{Fig-4}(b)]. The experimental results clearly show that there exists straight DNL in the path of SR. The experimentally measured frequency range of the DNL appears to be wide due to the limited sample size. The experimental results also observed the cross point generated by the hourglass WNL along the path of $\Gamma$Y.

Next, we experimentally investigate the surface responses of the (010) and (001) surfaces of the Dirac NLSMs.
The resin plate are fabricated on the (010) and (001) surfaces to mimic the hard-wall surface boundary used in full-wave simulations. The method of surface pump-probe spectroscopy is used to obtain the surface dispersions.
A subwavelength acoustic source is placed at the corresponding surface and then the detector scans the Bloch states near the sample surface. The measured surface dispersions of the (010) surface are shown in Fig.\ref{Fig-4}(c), which are in good agreement with the simulation results (see red dots). To show that the surface states on the (010) surface cover the entire Brillouin zone, we also present the results for $k_x=0.3\pi/p_x$, as shown in Fig.\ref{Fig-4}(d). The presence of surface states clearly confirms this [see Fig.\ref{Fig-4}(d)]. The isofrequency contours of the (010) surface for different frequencies are shown in Fig.\ref{Fig-4}(e-j). When the measurement frequency is lower than the DNL frequency range, the starting point of the topological surface arc starts from the boundary of the Brillouin zone and ends at the bulk state [The gray regions in Fig.\ref{Fig-4}(e-g) are the bulk band projections.], see red curves in Fig.\ref{Fig-4}(e-g).  The topological surface arc is an open profile and its starting and ending position will change with measurement frequency, shown in Fig.\ref{Fig-4}(e-g).
However, the topological surface arc is closed profile when the measurement frequency is larger than the DNL frequency range, as shown in Fig.\ref{Fig-4}(h-i). When the measurement frequency exceeds a critical value, the topological surface states disappear, see Fig.\ref{Fig-4}(j). Therefore, surface response of the DNL differs from Weyl semimetals where the Fermi arcs are always open profiles. The measured surface dispersions of the (001) surface are shown in Fig.\ref{Fig-4}(k), which clearly show that the topological state occupies only part of the Brillouin zone.

\textcolor{blue}{\textit{Conclusion}}. In summary, we have theoretically and experimentally realized, for the first time,  the long-sought Dirac nodal line semimetals. We further demonstrate that the Berry phase of Dirac nodal lines is $2\pi$, which is completely different from that of Weyl nodal lines with $\pi$ Berry phase. This also leads to a completely different surface response for parallel Dirac nodal lines than conventional Weyl NLSMs. For the designed acoustic crystal, the (100) plane has no any surface states, while the surface states can cover the whole Brillouin zone for the (010) plane and thereby the novel torus surface states are realized. Meanwhile, this system also has drumhead-like surface states caused by the hourglass Weyl nodal line. Further, we reveal that topological surface arcs change from open to closed contours on the (010) plane, which is very different from that of Weyl semimetals. The versatile surface response of the Dirac NLSMs provides an ideal platform for manipulating sound waves, for example surface acoustic wave isolator and  omnidirectional surface acoustic wave device.

\textcolor{blue}{\textit{Acknowledgements}}. This work is supported by the National Natural Science Foundation of China (Grant Nos. 11734003, 61975016, 12061131002, 12102039, 12204041, 12272040, 92163206); the Strategic Priority Research Program of Chinese Academy of Sciences (Grant No. XDB30000000); the National Key R\&D Program of
China (Grant Nos. 2020YFA0308800, 2021YFA1400100); Natural Science Foundation of Beijing Municipality (Grant Nos. Z190006 and 1212013); China Postdoctoral Science Foundation (Grant No. 2021M700436).


\end{document}